\documentclass{emulateapj} 

\usepackage{natbib,threeparttable,apjfonts,amsmath,booktabs,color,multirow,ulem}

\newcommand{\xmm}{{\it XMM--Newton}}
\newcommand{\fermi}{{\it Fermi}-LAT}

\newcommand{\nH}{$N_{\text{H}}$}
\newcommand{\snr}{CTB~109}
\newcommand{\ps}{1E 2259+586}
\newcommand{\mosi}{MOS~1}
\newcommand{\mosii}{MOS~2}
\newcommand{\mosiii}{MOS~1/2}

\shortauthors{Castro et al.}
\slugcomment{Accepted for publication in  ApJ}

\begin{document} 
\title{{\it Fermi}-LAT Observations and A Broadband Study of Supernova Remnant CTB 109}
\author{Daniel Castro\altaffilmark{1}, Patrick Slane\altaffilmark{2}, Donald C. Ellison\altaffilmark{3}, and Daniel J. Patnaude\altaffilmark{2}}

\altaffiltext{1}{MIT-Kavli Center for Astrophysics and Space Research, 77 Massachusetts Avenue, Cambridge, MA, 02139, USA}
\altaffiltext{2}{Harvard-Smithsonian Center for Astrophysics, 60 Garden Street, Cambridge, MA 02138, USA}
\altaffiltext{3}{Physics Department, North Carolina State University, Box 8202, Raleigh, NC 27695, USA}

\begin{abstract}

\snr\ (G109.1-1.0) is a Galactic supernova remnant (SNR) with a hemispherical shell morphology in X-rays and in the radio band. In this work we report the detection of $\gamma$-ray emission coincident with \snr, using 37 months of data from the Large Area Telescope on board the {\it Fermi} Gamma-ray Space Telescope. We study the broadband characteristics of the remnant using a model that includes hydrodynamics, efficient cosmic ray acceleration, nonthermal emission and a self-consistent calculation of the X-ray thermal emission. We find that the observations can be successfully fit with two 
distinct parameter sets, one where the $\gamma$-ray emission is produced 
primarily by leptons accelerated at the SNR forward shock and the other 
where $\gamma$-rays produced by forward shock accelerated cosmic-ray 
ions dominate the high-energy emission. Consideration of thermal X-ray emission introduces a novel element to the broadband fitting process, and while it does not rule out either the leptonic or the hadronic scenarios, it constrains the parameter sets required by the model to fit the observations. Moreover, the model which best fits the thermal and nonthermal emission observations is an intermediate case, where both radiation from accelerated electrons and hadrons contribute almost equally to the $\gamma$-ray flux observed.

\end{abstract}

\keywords{acceleration of particles --- cosmic rays --- gamma rays: ISM  --- ISM: individual (CTB 109) --- ISM: supernova remnants}

\section{Introduction}
The origin of cosmic rays (CRs) with energies up to the {\it knee} of the CR energy spectrum is broadly attributed to the acceleration of particles at the shocks of supernova remnants (SNRs) in our Galaxy, as reviewed in \citet{Reynolds2008}. Observational evidence that suggests this is the case includes the detection of non-thermal X-ray emission from young shell-type SNRs, such as RX J1713.7-3946 \citep{Koyama1997,Slane1999}, and Vela Jr. \citep{Aschenbach1998,Slane2001}, since these X-rays are believed to be synchrotron radiation from electrons accelerated to TeV energies at the SNR shock through diffusive shock acceleration (DSA). Observations of $\gamma$-ray emission in the TeV range from some SNRs also support the scenario where particles are being accelerated to energies approaching 10$^{15}$ eV in these objects \citep{Muraishi2000,Aharonian2004,Katagiri2005}. However, it has proven difficult to ascertain whether these high energy photons result from leptonic emission mechanisms (such as inverse Compton or non-thermal bremsstrahlung emission), or from relativistic hadrons interacting with the ambient medium \citep[e.g.][]{Ellison2010,Inoue2012}. 

Cosmic rays interacting with regions of high-density material are expected to result in $\gamma$-ray emission from the decay of neutral pions generated by proton-proton interactions, as suggested by \citet{Claussen1997}. Hence, studying SNRs that appear to be propagating into molecular clouds (MCs) is particularly important when searching for signatures of CR acceleration in remnants. GeV-TeV observations have proven that SNR/MC systems are likely an important class of $\gamma$-ray sources, as supported by correlation studies carried out by \citet{Hewitt2009}. The Large Area Telescope (LAT), onboard the {\it Fermi Gamma-ray Telescope} has allowed successful detections of several such systems in the GeV energy range such as SNRs W51 \citep{Abdo2009c}, G349.7--0.5, CTB~37A, 3C~391, and G8.7--0.1 \citep{Castro2010}. Recently, observations of the SNR W44 in the MeV-GeV energy range with AGILE/GRID were combined with Fermi-LAT data and the characteristics of the resulting spectrum strongly suggest that the $\gamma$-ray emission originates from $\pi^0$-decay of CR hadrons interacting with ambient material \citep{Giuliani2011}. On the other hand, observations of SNR RX J1713.7-3946 with Fermi-LAT have been used to argue that its $\gamma$-rays are produced from leptonic processes \citep{Abdo2011aa,Ellison2010}. However, alternative scenarios where the SNR is expanding into complex surroundings, such as clumps of interstellar material, offer plausible explanations for the broadband characteristics of the remnant where the $\gamma$-rays result from pion decay emission \citep[e.g.][]{Inoue2012}.

In addition to the information gathered through observations of non-thermal radiation from SNRs, studies of the hydrodynamic evolution and the properties of the shocked medium and ejecta have yielded important information about CR production. The X-ray morphology of Tycho's SNR and SN~1006 indicates that the shock compression ratios in these objects have been modified by the acceleration of cosmic ray ions \citep{Warren2005,Cassam2008}, while comparisons of the postshock plasma temperatures and shock velocities observed in SNRs 1E~0102.2-7219 and RCW~86 also point to a significant fraction of their explosion energy being placed in relativistic particles \citep{Hughes2000,Helder2009}. 

The Galactic SNR \snr\ (G109.1-1.0) was originally discovered as an extended X-ray source by \citet{Gregory1980}, with the {\it Einstein} X-ray Observatory. The morphology of the source in the 0.1-4.5 keV energy range was identified as that of a semicircular shell, and corresponds well to that in the radio band, as observed in subsequent $\lambda$49 cm observations by Hughes et al. (1981, 1984), with the Westerbork Synthesis Radio Telescope (WSRT). Using \xmm\ observations, \citet{Sasaki2004} performed a detailed X-ray study of \snr, and found the emission from the remnant to be thermal in nature.

The environment into which \snr\ is expanding is a complex one, and is likely the source of the particular morphology of this remnant. There is a giant molecular cloud (GMC) located nearby, west of the SNR, which contains several H~{\scshape ii} regions, including the cloud S152 \citep{Israel1980}. From $^{12}$CO and $^{13}$CO observations \citet{Tatematsu1987} conclude that \snr\
is in contact with the GMC, and that the molecular material is responsible for the anisotropic SNR expansion. Since the SNR is a half shell, not only in the X-ray band, but also in radio continuum observations where the effects of absorption by the circumstellar and interstellar media are negligible, the lack of emission to the west is attributed to the SNR shock interacting with the cloud of dense material. \citet{Kothes2002} compare the results of H~{\scshape i} and CO observations towards \snr, with the parameters obtained for H~{\scshape ii} regions in the complex \citep{Tatematsu1990}, and derive a distance of $D_{\text{SNR}}=3.0\pm0.5$ kpc. Studies by \citet{Durant2006} and \citet{Tian2010} have found larger estimates of the distance to CTB 109 (7.5 kpc and 4 kpc, respectively). Recently, this debate has been thoroughly discussed in \citet{Kothes2012}, who determined the distance to be 3.2$\pm0.2$ kpc.

In addition to the complexities of its interaction with its surroundings, \snr\ is also of great interest because there is a magnetar located within it. \citet{Gregory1980} identified this compact source at approximately the center of curvature of the SNR shell, and it has since been determined to be an anomalous X-ray pulsar (AXP), \ps\ \citep[and references therein]{Woods2004}. No radio counterpart has been found to the X-ray pulsar however \citep{Kothes2002}. It has been suggested that the ejecta of SN explosions that result in magnetars as compact remnants of the progenitors should be more energetic than 10$^{51}$ ergs, the canonical value for other SN explosions \citep{Duncan1992,Bucciantini2007}. However, no evidence for explosion energies larger than 10$^{51}$ erg has been found in observations of other SNRs with magnetars, such as Kes~73 and N49  \citep{Vink2006}. \citet{Sasaki2004} determined the explosion energy of \snr\ to be $E=0.7 \times 10^{51} d_3$ erg,  where $d_3=D_{\text{SNR}}/(3\text{ kpc})$, using the X-ray observations of the remnant and assuming instantaneous equilibration of ion and electron temperatures. Since \citet{Durant2006} estimated the distance to \snr\ to be $D_{\text{SNR}}\approx7.5$ kpc using observations of nearby red clump stars, they conclude that the explosion energy must have been $\sim7\times10^{51}$, and therefore that this remnant is an example of the "super-energetic explosions" expected from SN where magnetars form. The most recent distance estimates appear to rule out this scenario.

Here we report on $\gamma$-ray observations of \snr\ with the \fermi, and we model the broadband characteristics of the remnant using a simulation that includes hydrodynamic evolution, DSA, nonthermal emission modeling and a self-consistent calculation of the X-ray thermal emission obtained by tracking the non-equilibrium ionization state of the shocked plasma \citep[and references therein]{Ellison2012,Patnaude2009,Patnaude2010}. We describe several distinct scenarios for which the broadband characteristics of \snr\ are well matched by the model using a distance of $\sim3$ kpc, where for some parameter sets the MeV-GeV emission observed originates in leptonic processes, others where $\pi^0$-decay emission is the dominant source, and a special scenario where a combination of leptonic and hadronic emission is required to fit the spectrum. We also briefly discuss the results of simulations of \snr\ assuming a larger distance of $7.5$ kpc, which requires the "super-energetic explosion" proposed by \citet{Durant2006}. The observations are described in Section 2, the model, fit parameters, and discussion of the results are presented in Section 3, and the our concluding remarks are included in Section 4.

\section{OBSERVATIONS OF SNR CTB 109}

\subsection{Fermi-LAT and CTB 109}

In this work, 37 months of data from the {\it Fermi Gamma-ray Space Telescope} Large Area Telescope ({\it Fermi}-LAT) (from August, 2008 until September 2011) are analyzed. Only events belonging to the Pass 7 V6 {\it Source} class, which reduces the residual background rate as explained in detail in \citet{Abdo2011p7}, have been selected for this study. The updated instrument response functions (IRFs) used are called ``Pass7 version 6'', which were developed using in-flight data \citep{Rando2009,Abdo2011p7}. Additionally, only events coming from zenith angles smaller than 100$^{\circ}$ are selected in order to reduce the contribution from terrestrial albedo $\gamma$-rays \citep{AbdoPRD}. All analyses include data from a circular region in the sky centered on the position of \snr, with radius 15$^\circ$.

\begin{figure}
\includegraphics[width=\columnwidth]{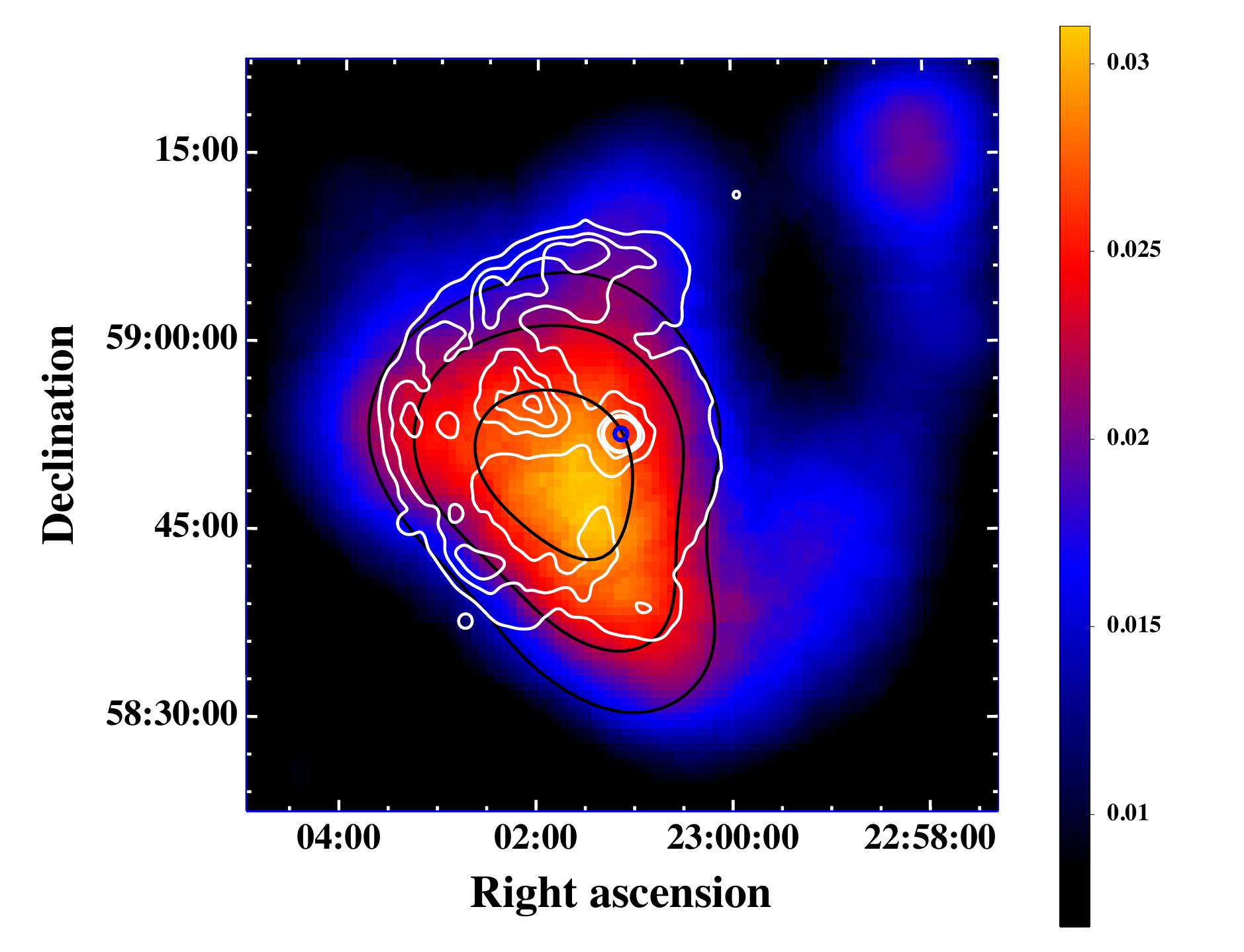}
\caption{Smoothed {\it Fermi} LAT count maps of {\it front} converted events in the range 2 to 200 GeV (units are $10^4 \text{counts deg}^{-2}$), of the $1^{\circ}\times1^{\circ}$ region, centered on SNR \snr. The pixel binning is 0.01$^\circ$, and the maps are smoothed with Gaussians of width 0.2$^\circ$. White contours represent the X-ray emission (0.5-5.0 keV) from \xmm\ data, as explained in Section 3. The black contour curves delimit region where TS$\geq25-36-49$. The small blue circle indicates the position of \ps.}
\label{fig:LATcm}
\end{figure}

The  $\gamma$-ray data in the direction of \snr\ are analyzed using the Fermi Science Tools v9r23p1\footnote{The Science Tools package and related documentation are distributed by the Fermi Science Support Center at http://fermi.gsfc.nasa.gov/ssc}. The maximum likelihood fitting technique is employed, using {\it gtlike}, to obtain morphological and spectral information \citep{Mattox1996}. The emission models used in {\it gtlike} include a Galactic diffuse component resulting from the interactions of cosmic rays  with interstellar material and photons, and an isotropic one that accounts for the extragalactic diffuse and residual backgrounds. In this work, the mapcube file {\tt gal\_2yearp7v6\_v0.fits} is used to describe the $\gamma$-ray emission from the Milky Way, and the isotropic component is modeled using the {\tt iso\_p7v6source.txt} table. 

In order to understand the spatial characteristics of the $\gamma$-ray emission in the field of \snr, data in the energy range 2 to 200 GeV, and converted in the {\it front} section, are used. The 68\% containment radius angle for normal incidence {\it front}-selected photons in this energy band is $\leq 0.3^\circ$. Galactic and isotropic backgrounds are modeled and test statistic (TS) maps are constructed using {\it gttsmap} to allow for detection significance estimates, and to best evaluate the position and possible extent of the source.  The test statistic is the logarithmic ratio of the likelihood of a point source being at a given position in a grid, to the likelihood of the model without the additional source, $2\text{log}(L_{\text{ps}}/L_{\text{null}})$. 

Figure \ref{fig:LATcm} presents a count map of {\it front} converted events in the range 2 to 200 GeV, of 1$^{\circ}\times1^{\circ}$ around \snr. The X-ray morphology of the remnant is presented as white contours (from \xmm\ data, as discussed in Section 3), and the black contour lines encircle the region where TS$>25-36-49$, equivalent to a detection significance of 5$\sigma$, 6$\sigma$, and 7$\sigma$ respectively. A $\gamma$-ray source is coincident with the position of the remnant with centroid $(\alpha_{\text{2000}},\delta_{\text{2000}}=23^{\text{h}}01^{\text{m}}48^{\text{s}}, +58^{\circ}49'48'')$, with $95\%$ confidence radius $=4'.2$. The significance of the detection, obtained from the evaluation of the peak of the TS map, is $\sim 7.8\sigma$.  The residual test statistic map, built by modeling a point source at the best-fit centroid of emission, shows no evidence that the source is spatially extended, since the residual TS values do not exceed 3.7$\sigma$ within 1$^\circ$ of the centroid of emission. The possible extension of this source was further explored by comparing the overall likelihood obtained with {\it gtlike} for spatial models of the emission from \snr\ as disk templates spanning several sizes ($r_{\text{disk}}=6'-12'-18'-24'$), as well as a spatial template of the X-ray emitting material as observed by \xmm\ (the X-ray observations are explained in detail in \S 2.2), with the likelihood of the point source model. The disk templates are moderately more successful at fitting the observations than the point source model ($\sim 40$), with TS peaking at $\sim 45$ for the $12'$ disk. The disk template with radius $18'$ is similarly successful (TS$\sim43$), butt the TS comparison suggests the morphology of the $\gamma$-ray emission from \snr\ is not consistent with that of the X-ray emission. Hence, we conclude that while we cannot rule out a point-like emission model, the morphology of the $\gamma$-ray source is possibly as extended as the SNR. However, the spatial distribution of emission is inconsistent with that of the thermal X-ray picture.

The study of the spectral energy distribution (SED) characteristics of the source associated with CTB 109, is performed using events converted in both {\it front} and {\it back} sections, and in the energy range 0.2-204.8 GeV. The lower energy bound is selected to avoid the rapidly changing effective area of the instrument at low energies, and because of the large uncertainty below 0.2 GeV related to the Galactic diffuse model used. {\it gtlike} is used to model the flux at each energy bin and estimate, through the maximum likelihood technique, the best-fit parameters. Background sources from the 24-month {\it Fermi} LAT Second Source Catalog \citep{Abdo2011b}\footnote{ The data for the 1873 sources in the {\it Fermi} LAT Second Source Catalog is made available by the Fermi Science Support Center at http://fermi.gsfc.nasa.gov/ssc/data/access/lat/2yr\_catalog/} have been included in the model likelihood fits. The systematic uncertainties of the effective area, for the IRF used, are energy dependent: 10\% at 100 MeV, decreasing to 5\% at 560 MeV, and increasing to 20\% at 10 GeV \citep[and references therein]{Porter2009,Abdo2011b}. As an addition to the statistical uncertainties associated with the likelihood approach, and the systematic errors related to the IRFs, the uncertainty of the underlying Galactic diffuse level is considered. This source of uncertainty is included in the evaluation of the systematic uncertainties by artificially varying the normalization of the Galactic background by $\pm6\%$ from the best-fit value at each energy bin, similarly to treatments presented in 
\citep{Castro2010}.

\begin{figure}
\includegraphics[width=\columnwidth]{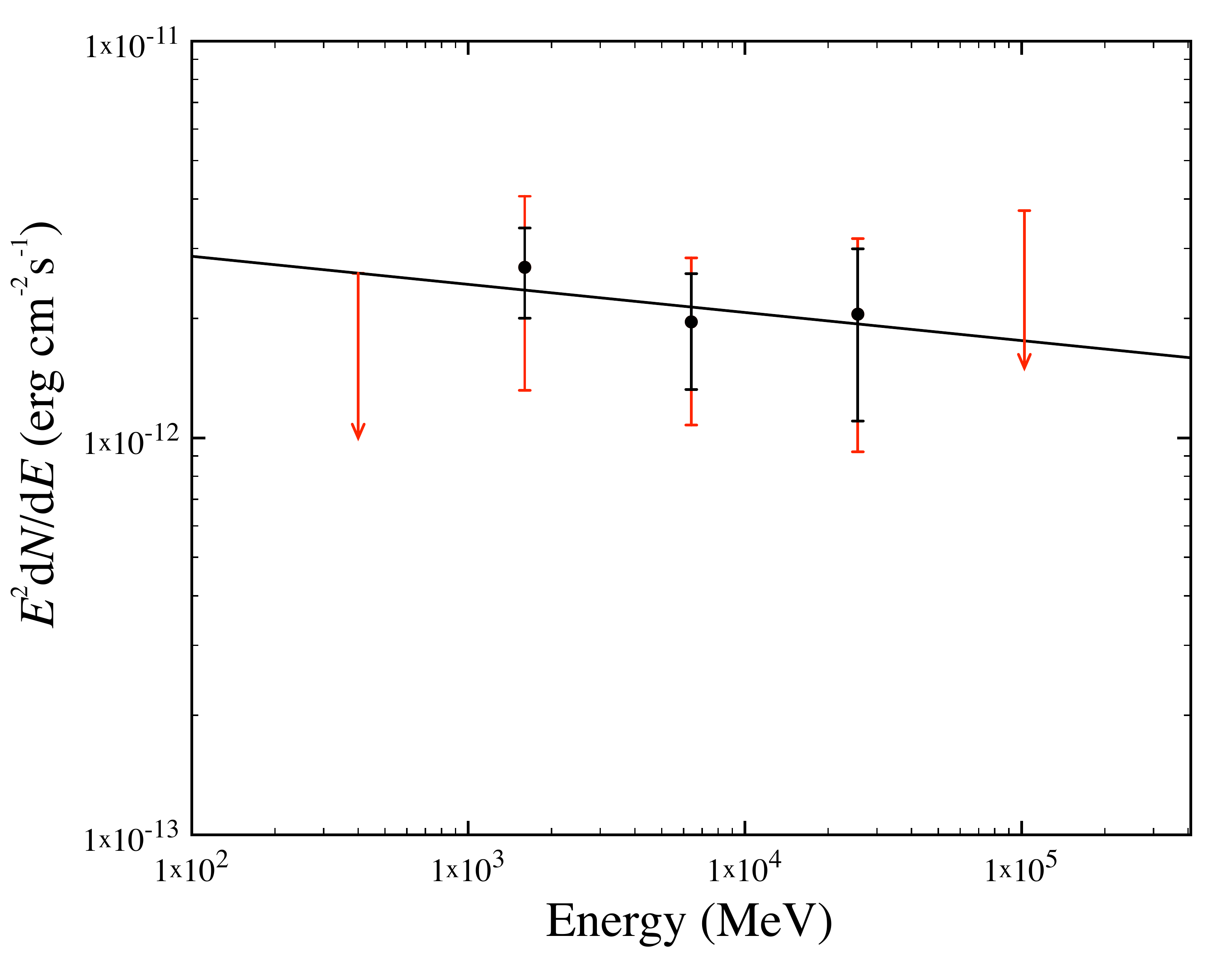}
\caption{{\it Fermi} LAT spectral energy distribution of the source coincident with SNR \snr. Statistical uncertainties are shown as black error bars, and systematic errors are indicated by red bars. The solid line represent the best-fit powerlaw model to the data.}
\label{fig:LATspec}
\end{figure}

The spectrum of the $\gamma$-ray emission is shown in Figure \ref{fig:LATspec}, where the best-fit power law model to the data is also presented. For energies below 800 MeV, and above 51.2 GeV only flux upper limits are determined from the data. The best-fit model yields a spectral index of $\Gamma = Ð2.07 \pm 0.12$, and an integrated photon flux above 100 MeV of $F_{>100\text{ MeV}}\approx 1.7\times10^{-8}$ photons/cm$^{2}$/s. 

\subsection{XMM-Newton Study}

\snr\ was observed with \xmm\ during period A01, when the entire remnant was covered with three pointings, distributed South, North and East (ObsIDs 00575401, 00575402, and 00575403, respectively). A detailed spatial and spectral analysis of these observations was previously presented in \citet{Sasaki2004}. We have reanalized the data in order to best adapt to the requirements of broadband modeling. Data from the European Photon Imaging Cameras (EPIC) \mosi, and \mosii\ are studied using the \xmm\ Science Analysis System (SAS) version 11.0.0\footnote{The SAS package, and related documentation, is distributed by the \xmm\ Science Operations Center at http://xmm.esac.esa.int/sas/}, starting from the observational data files (ODFs). Filtering and data reduction details follow those in \citet{Castro2011a}, and references therein. The effective exposure time after standard screening is 14.1 (14.5) ks for \mosi\ (\mosii). The EPIC PN camera observations were heavily affected by cosmic ray noise and bad pixel columns, hence making the PN data difficult to match to the regions studied with the MOS 1/2 cameras which have better energy resolution. The PN observations are not used in this study.

\begin{figure}
\includegraphics[width=\columnwidth]{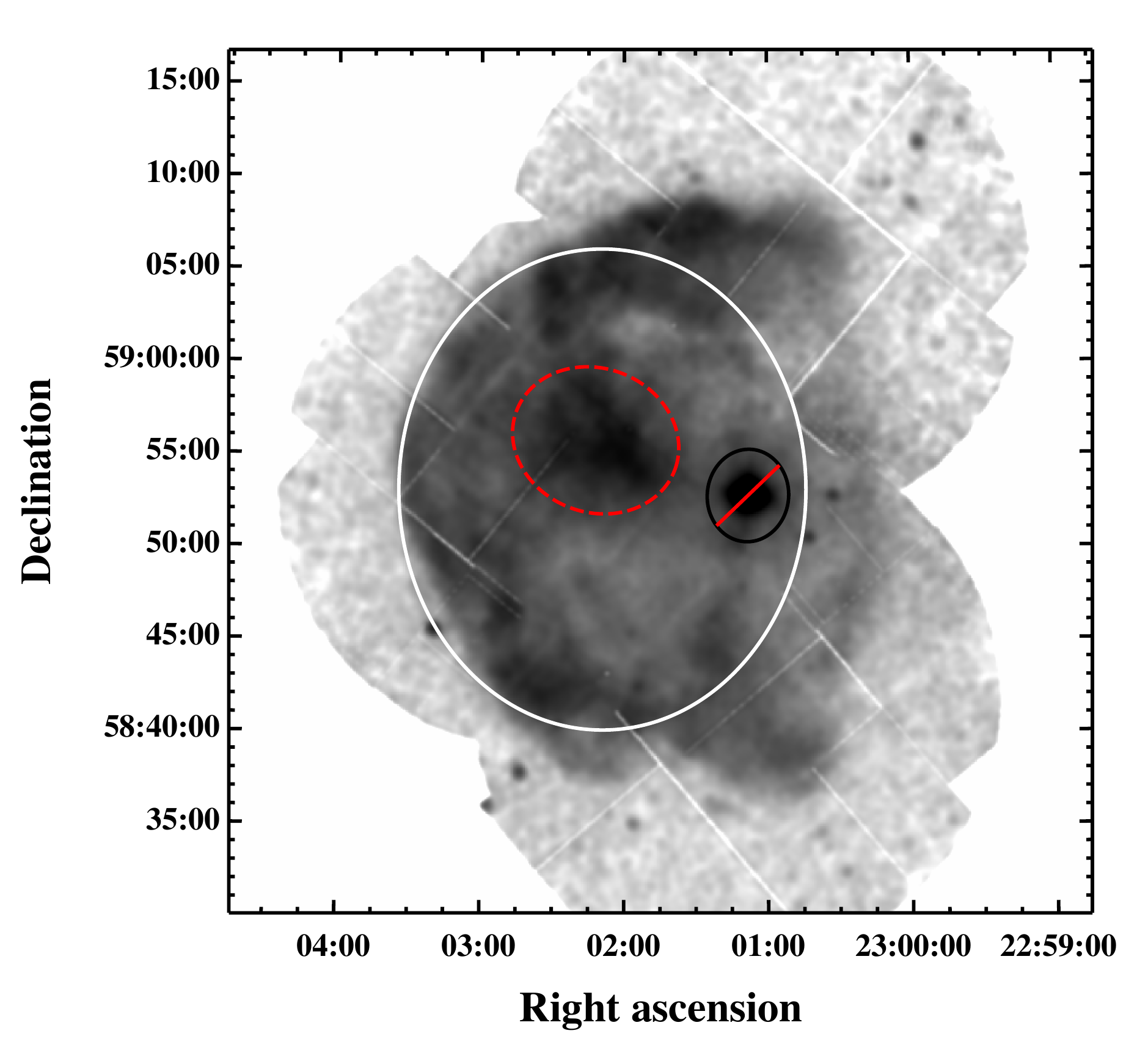}
\caption{Smoothed, merged \xmm\ MOS 1/2 intensity map in the range 0.5-5.0 keV, of the three observations towards \snr. The white ellipse indicates the region selected for spectrum extraction. The black ellipse represents the area associated with emission from the AXP \ps, which has been removed from the analysis, and the red (dashed) ellipse shows the approximate extent of the X-ray emission enhancement known as the ``Lobe'' \citep[and references therein]{Sasaki2004}.}
\label{fig:XMMmap}
\end{figure}

The \xmm\ EPIC MOS mosaic image of \snr, in the 0.5-5.0 keV band, is presented in Figure \ref{fig:XMMmap}. To investigate the spectral characteristics of the SNR, we extracted the spectrum for the region shown as a white ellipse, which covers approximately 45\% of the extent of the SNR, if we project its hemispheric X-ray morphology onto a full shell, with radius 18$'$.5 (estimated from the distance between the \ps\ and the shock position towards the East). Emission from the AXP \ps\ was not included, nor was emission from other point sources in the field. Corresponding effective area and spectral response files for the extended extraction region were created, and the ancillary response file obtained allowed for vignetting effects to be corrected. In order to account for the total \xmm\ background, a two step process is performed which uses both corrected blank sky products provided by the \xmm\ EPIC Background Working Group\footnote{The blank sky data, obtained using observations of several sky positions, is a combination of the Cosmic X-ray background and the non X-ray background, which varies in time and results from a combination of the interaction of cosmic rays with the detector, background due to solar protons, and electronic noise \citep{Carter2007}. Blank sky data from the \xmm\ EPIC Background Working Group are available at http://xmm2.esac.esa.int/external/xmm\_sw\_cal/background/.}, and data from a region outside the SNR shell, similar to the method detailed in \citet{Arnaud2002}. 

\begin{figure}
\includegraphics[width=\columnwidth]{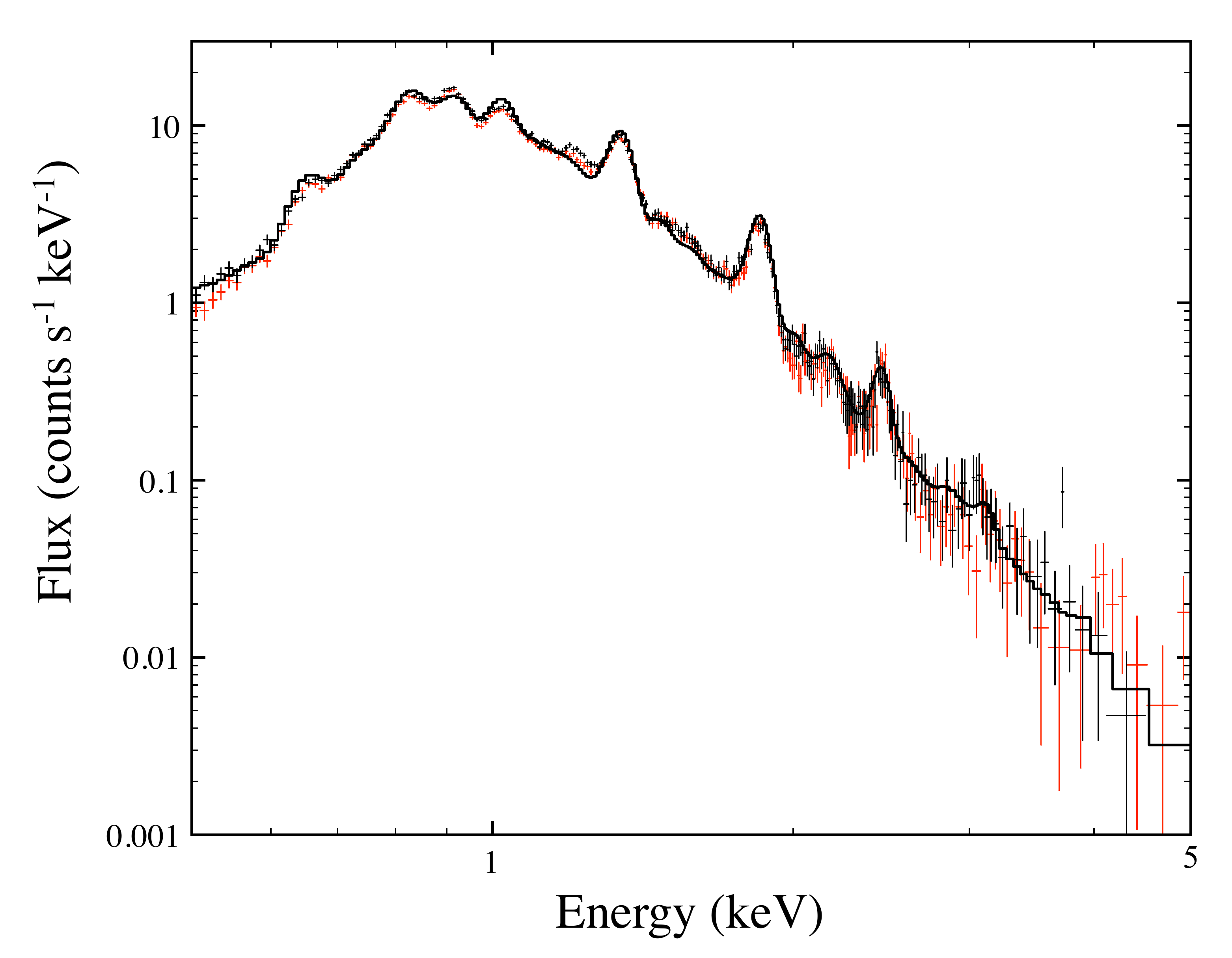}
\caption{\xmm\ EPIC MOS 1/2 background subtracted spectra of the region illustrated in Figure \ref{fig:XMMmap} of \snr. The best-fit {\scshape tbabs*sedov} model is shown as a histogram, and its parameters are presented in Table \ref{tab:mos}}
\label{fig:XMMspec}
\end{figure}

The resulting \mosiii\ spectra were analyzed with the {\scshape xspec} software package (version 12.5.1), using an energy range of 0.4-5.0 keV, grouped with a minimum of 25 counts bin$^{-1}$, and using the $\chi^2$ statistic (see Figure \ref{fig:XMMspec}). The emission above $\sim 5$ keV is dominated by the background. The model for the absorbed emission from an SNR in the Sedov-Taylor expansion phase {\scshape sedov} (NEIVERS 2.0) describes well the spectral characteristics \citep{Borko2001}. The best-fit parameters for this model are shown in Table \ref{tab:mos}, where the uncertainties quoted are the 90\% confidence limits (1.6$\sigma$). In this work, the T\"ubingen-Boulder model for the absorption of X-rays in the interstellar medium (ISM) {\scshape tbabs} is adopted \citep{Wilms2000}.

\begin{table}
\begin{center}
\begin{threeparttable} 
\caption{Spectral Fit Parameters from \mosiii\ data with {\scshape sedov} model}
\label{tab:mos}
\begin{tabular}{lcc}
\toprule
Parameter&&Value \\
\midrule
Column Density, $N_{\text{H}}$ ($10^{22} \text{cm}^{-2}$)&&0.67$^{+0.01}_{-0.01}$\\
\noalign{\smallskip}
Mean Shock Temperature, $kT_{\text{S}}$ (keV)&&0.38$^{+0.03}_{-0.03}$\\
\noalign{\smallskip}
Electron Temperature \tablenotemark{a}, $kT_{\text{e}}$ (keV)&&0.18$^{+0.03}_{-0.03}$\\
\noalign{\smallskip}
Ionization Timescale, $n_{\text{e}}t$ ($\times 10^{12}$\,s cm$^{-3}$)&&1.27$^{+0.07}_{-0.05}$\\
\noalign{\smallskip}
Absorbed X-ray Flux \tablenotemark{b}, $F_{\text{abs}}$ (erg cm$^{-2}$ s$^{-1}$)&&$6.7\times10^{-11}$\\
\noalign{\smallskip}
Unabsorbed X-ray Flux \tablenotemark{c}, $F_{\text{unab}}$ (erg cm$^{-2}$ s$^{-1}$)&&$7.9\times10^{-10}$\\
\noalign{\smallskip}
Reduced $\chi^2$ Statistic&&3.4(499 dof)\\
\noalign{\smallskip}
\bottomrule
\noalign{\smallskip}
\end{tabular}

\begin{tablenotes}[para]
\item[a]{$kT_{\text{e}}$ is the electron temperature immediately behind the shock.}

\item[b]{$F_{\text{abs}}$ is the average absorbed X-ray flux of the region, calculated in the 0.5-5.0 keV energy band, using the best-fit model to the EPIC \mosiii\ data.}

\item[c]{$F_{\text{unab}}$ is the average unabsorbed X-ray flux of the region, calculated in the 0.5-5.0 keV energy band, using the best-fit model to the EPIC \mosiii\ data.}
\end{tablenotes}

\end{threeparttable} 
\end{center}
\end{table}

\subsection{Radio Observations}

As part of the Canadian Galactic Plane Survey (CGPS), SNR \snr\ was studied with the Dominion Radio Astrophysical Observatory (DRAO, \citealt{Kothes2006}). The semi-circular morphology is consistent with previous observations in the radio band by Hughes et al. (1981,1984), as well as the spectral characteristics: index $\alpha=0.5$, where the radio flux is modeled as $S_\nu \propto \nu^{-\alpha}$, and total flux at 1.4 GHz of approximately $17$ Jy. The radio spectral data used in Section 4 is a combination of the information obtained by  \citet{Hughes1984} and \citet{Kothes2006}.

The CO line data used to estimate the location of molecular material around \snr\ in Section 3.5 was also originally gathered by the CGPS, from observations of $^{13}$CO (J=1-0) emission with the Five College Radio Astronomy Observatory (FCRAO) at 115.3 GHz \citep{Taylor2003}. 

\section{Analysis and Discussion}

\subsection{Hydrodynamic Evolution and Broadband Emission Model}

The simulations presented here are calculated with CR-hydro-NEI, a code that self-consistently includes the SNR hydrodynamics, a semi-analytic calculation of nonlinear DSA, the nonequilibrium ionization conditions behind the forward shock, the broadband continuum from radio to gamma-rays, and the X-ray line emission from the interaction region between the contact discontinuity and the forward shock. The X-ray emission profile is calculated using the NEI information for the thermal radiation component along with the X-ray synchrotron emission from relativistic electrons.The CR-hydro-NEI model is described in greater detail in \citet{Ellison2007}, Patnaude et al. (2009,2010), and references therein. This model also includes CR escape from the shock, and how this population of accelerated particles diffuses into density enhancements upstream. A full description of how this is implemented is included in \citet{Ellison2011}\footnote{We note that recently an extensive generalization of the 
CR-hydro-NEI code has been 
published \citep[e.g.,][]{2012ApJ...750..156L}. The generalized version includes a number of effects that are not implemented in the version used here. These include (1) an explicit calculation of the upstream precursor structure, (2) a momentum and space dependent CR diffusion coefficient, (3) an explicit calculation of MFA replacing the ad hoc parameterization of MFA used here, and (4) a finite Alfven speed for the particle scattering centers based on the amplified magnetic field. These generalizations are important for modeling certain plasma physical processes associated with NL DSA, but they involve a number of new parameters that are not yet strongly constrained by observations such as those available for CTB109. 
\citet{2012ApJ...750..156L} showed for SNR RX J1713, where 
high-quality, broadband data exists, that models using the generalized CR-hydro-NEI code give essentially idenical fits to the broadband continuum and X-ray emission line data as the earlier version of 
CR-hydro-NEI used here with a modest change in parameters. Except for this modest change in parameters, all of the conclusions reached here are unchanged in the generalized version of CR-hydro-NEI and we have chosen to use the simpler model described in Ellison et al. (2012) for modeling CTB109. 
}.

Through the evolution of the simulation, the NEI state of the shocked plasma, between the forward shock (FS) and the contact discontinuity, is calculated self-consistently at each time step, using the ionization structure, free electron number density, and electron temperature. The \citet{Raymond1977} plasma emissivity code, updated by \citet{Brickhouse1995} is then used to obtain the thermal X-ray emission from the system, for the resulting plasma characteristics (Patnaude et al. 2010). 

The electron spectrum from the acceleration process is used to calculate the synchrotron and nonthermal bremsstrahlung radiation from the system, as well as the inverse Compton (IC) emission assuming standard seed photons from the cosmic microwave background. The model also considers the nonthermal hadronic emission process by calculating the pion decay emission spectrum \citep{Kamae2006}, resulting from collisions between protons accelerated to relativistic speeds at the FS and those in the surrounding material. 

\subsection{Model Parameters}

The distance of $D_{\text{SNR}}=3.0\pm0.5$ pc, estimated by \citet{Kothes2002} from CO and H~{\scshape i} observations, is adopted, and only values within the uncertainty range are considered. Given the apparent angular size of the remnant from the X-ray image, the radius of the semicircle is estimated to be $R_{\text{SNR}}=18'\pm1'=(16\pm1)d_3$ pc, where $d_3=D_{\text{SNR}}/(3\text{ kpc})$, similar to the analysis in \citet{Sasaki2004}. As previously mentioned, the X-ray and radio morphology of \snr\ and its apparent vicinity to the GMC suggest the remnant expanded into the dense molecular material towards the west and rapidly decelerated in that direction, resulting in its semicircular shape. This scenario is consistent with the results obtained by \citet{Wang1992} using hydrodynamic modeling of this remnant evolving at the interface of the diffuse interstellar medium and a dense ($n_{\text{cloud}}=36$ cm$^{-3}$) molecular cloud region. Since CR-Hydro-NEI is a spherically symmetric model, the simulations have been allowed to evolve to a given radius (dictated by the angular size, and the distance) and the resulting model fluxes for fitting have been scaled to appropriately account for the observed emission originating in only a fraction of the simulated volume. Hence, the age of the remnant, $t_{\text{SNR}}$, is varied in the model for the different parameter sets to match the radius, $R_{\text{SNR}}$. 

While the likely association with AXP \ps\ strongly suggests SNR \snr\ is the result of a core-collapse supernova (SN), we follow the majority of work on this remnant and explore cases where the ambient density profile is that of a uniform circumstellar environment with constant pre-shock proton number density $n_{\text{0}}$, and magnetic field $B_0$. \citet{Kothes2002} found that the structure of the H {\scshape i} in the region of \snr\ indicates that the SNR is not expanding into a stellar wind density profile, and hence suggest that the progenitor was a star of type B2/3. Additionally, the large swept up mass calculated from the X-ray observations of the remnant ($M_{\text{sw}}\sim$100 $M_{\odot}$, \citealt{Sasaki2004}) indicates that the emission and evolution of the SNR must be dominated by material outside any possible stellar wind from the progenitor.

 
A power-law ejecta density profile is assumed, as expected for core-collapse supernovae, with index 7 and ejecta mass, $M_{\text{ej}}=5$~M$_{\odot}$. Small variations of the characteristics of the ejecta, i.e., different ejecta mass values and power-law distribution indexes, were found to have negligible impact on the results of the simulations presented here. This is likely an effect of the advanced evolutionary stage of the modeled SNR, when the overall properties of the SNR are governed by the interaction between the shock and the CSM, rather than the characteristics of the ejecta. The analysis by \citet{Sasaki2004} suggests the swept-up mass of circumstellar material is large, as mentioned previously, and hence the contribution from the reverse shocked ejecta to the X-ray flux is likely modest and is not included in the models. 

\begin{table}
\begin{center}
\begin{threeparttable} 
\caption{CR-Hydro-NEI Input Parameters}
\label{tab:crhydro}
\begin{tabular}{lcc}
\toprule
Parameter&&Value \\
\midrule
SN Explosion Energy, $E_{\text{SN}}$ ($10^{51} \text{erg}$)&&1\\
\noalign{\smallskip}
Total Ejecta Mass, $M_{\text{ej}}$ ($M_{\odot}$)&&5\\
\noalign{\smallskip}
Pre-shock Proton Temperature, $T_0$ ($10^3$ K)&&10\\
\noalign{\smallskip}
Pre-shock Magnetic Field , $B_0$ ($\mu$G)&&4.5\\
\noalign{\smallskip}
Temperature of CMB Photon field , $T_{\text{CMB}}$ (K)&&2.725\\
\noalign{\smallskip}
\bottomrule
\noalign{\smallskip}
\end{tabular}

\end{threeparttable} 
\end{center}
\end{table}

Several other parameters characterizing the SN progenitor and the surrounding environment are required, including the SN explosion energy, $E_{\text{SN}}$ and the average temperature of the pre-shock circumstellar medium (CSM), $T_0$. Since magnetic field amplification (MFA) appears to be a key effect of particle acceleration in young SNR shocks (e.g., \citealt{Uchiyama2007}), an ad hoc magnification factor, $B_{\text{amp}}$, is included as an input parameter in the model. 
Some of the key model parameters common to all scenarios studied are presented in Table \ref{tab:crhydro}, while the age, number density of ambient protons, relativistic electron to relativistic proton ratio, magnetic amplification factor, and acceleration efficiency, $\varepsilon_{\text{CR}}$, have been varied in order for the models to appropriately fit the observed radius of the remnant, the broadband nonthermal spectrum (both in the radio band and $\gamma$-rays), and the thermal X-ray emission. Detailed descriptions and discussions of the input parameters required for the CR-Hydro-NEI model are provided in \citet{Ellison2007}, \citet{Patnaude2009}, and references therein.
This work focuses on several sets of parameters: those where the $\gamma$-rays originate in pion decay emission from cosmic ray ions produced in the SNR interacting with the ambient medium (the ``hadronic'' case); those for which the \fermi\ spectrum is the result of inverse Compton emission from accelerated electrons (the ``leptonic'' scenario); and mixed scenarios, where the leptonic and hadronic mechanisms both significantly contribute to the $\gamma$-ray flux. While the dominant source of $\gamma$-rays in the ``leptonic'' case is emission from accelerated electrons, all models place most of the accelerated particle energy in protons.

\subsection{Uniform Circumstellar Medium Density Model}

\begin{figure*}
\begin{center}
\includegraphics[width=0.32\textwidth]{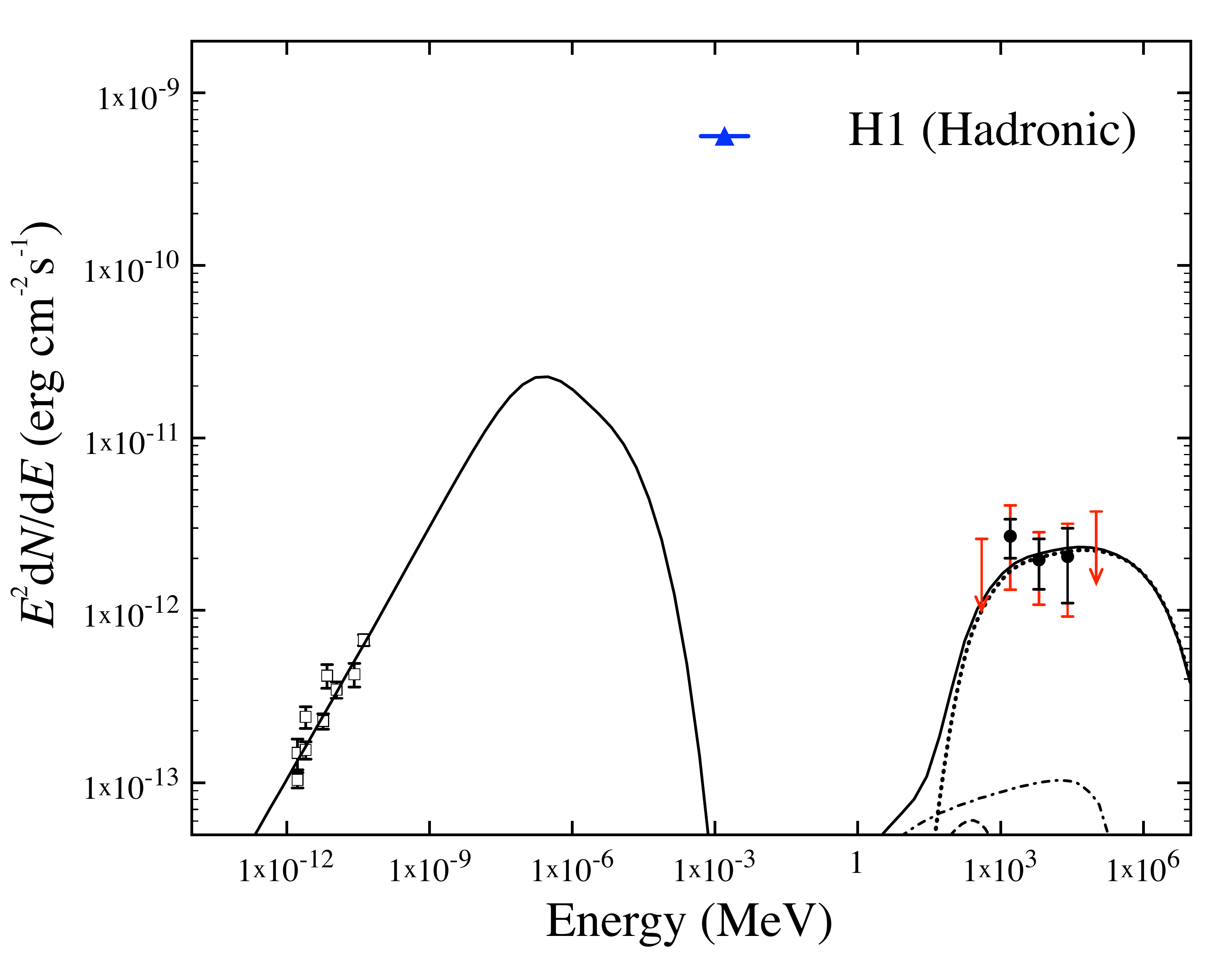}
\includegraphics[width=0.32\textwidth]{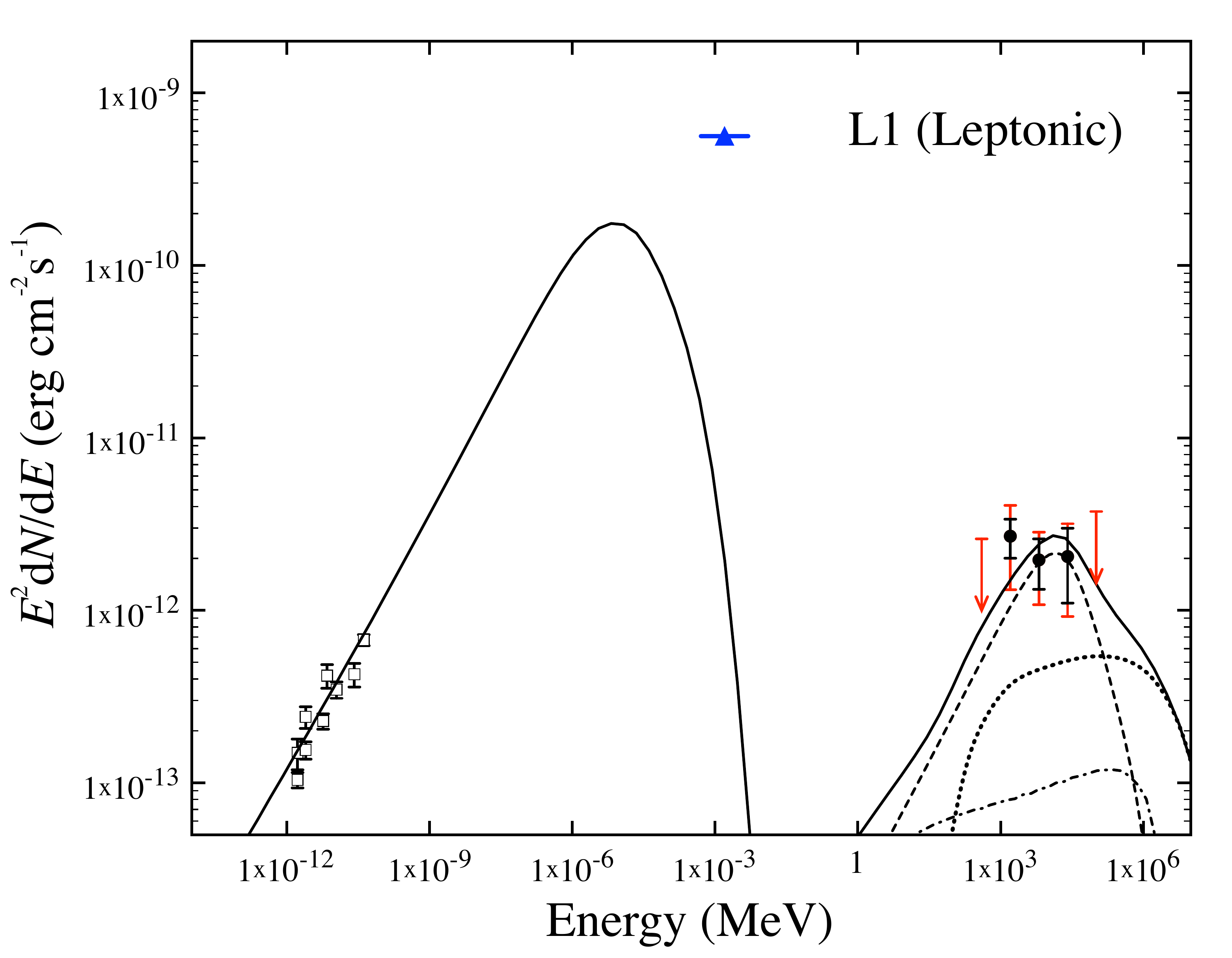}
\includegraphics[width=0.32\textwidth]{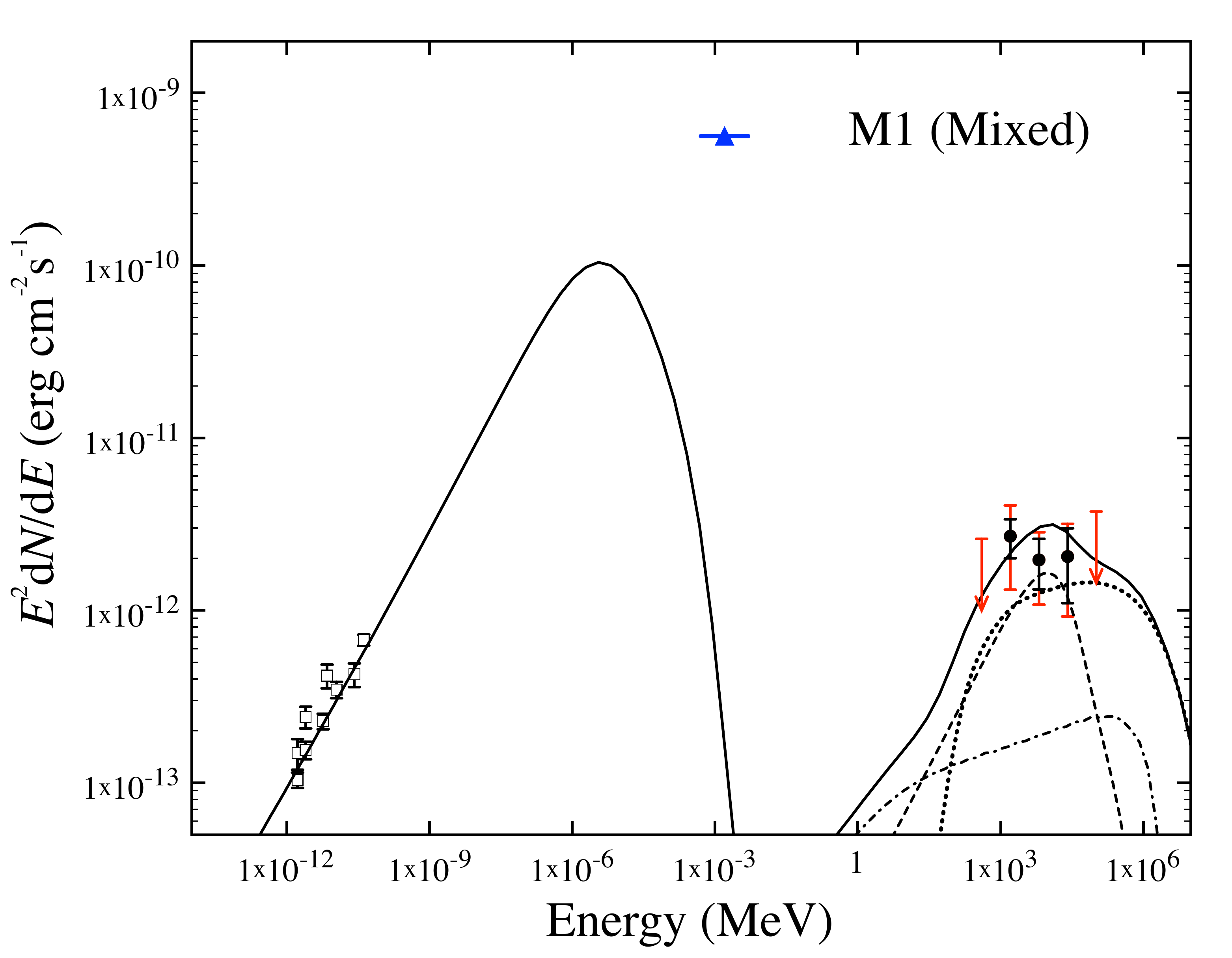}\\
\includegraphics[width=0.32\textwidth]{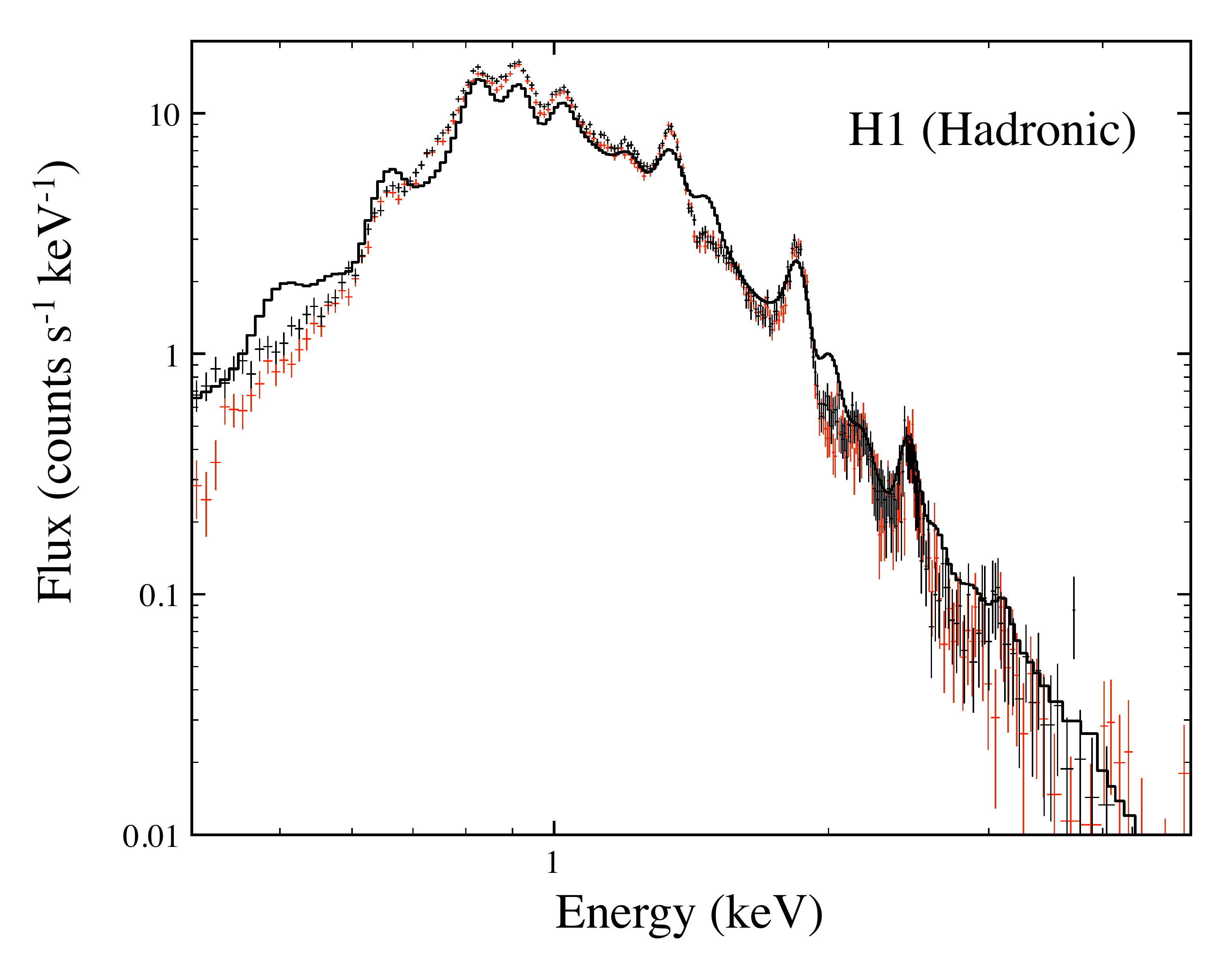}
\includegraphics[width=0.32\textwidth]{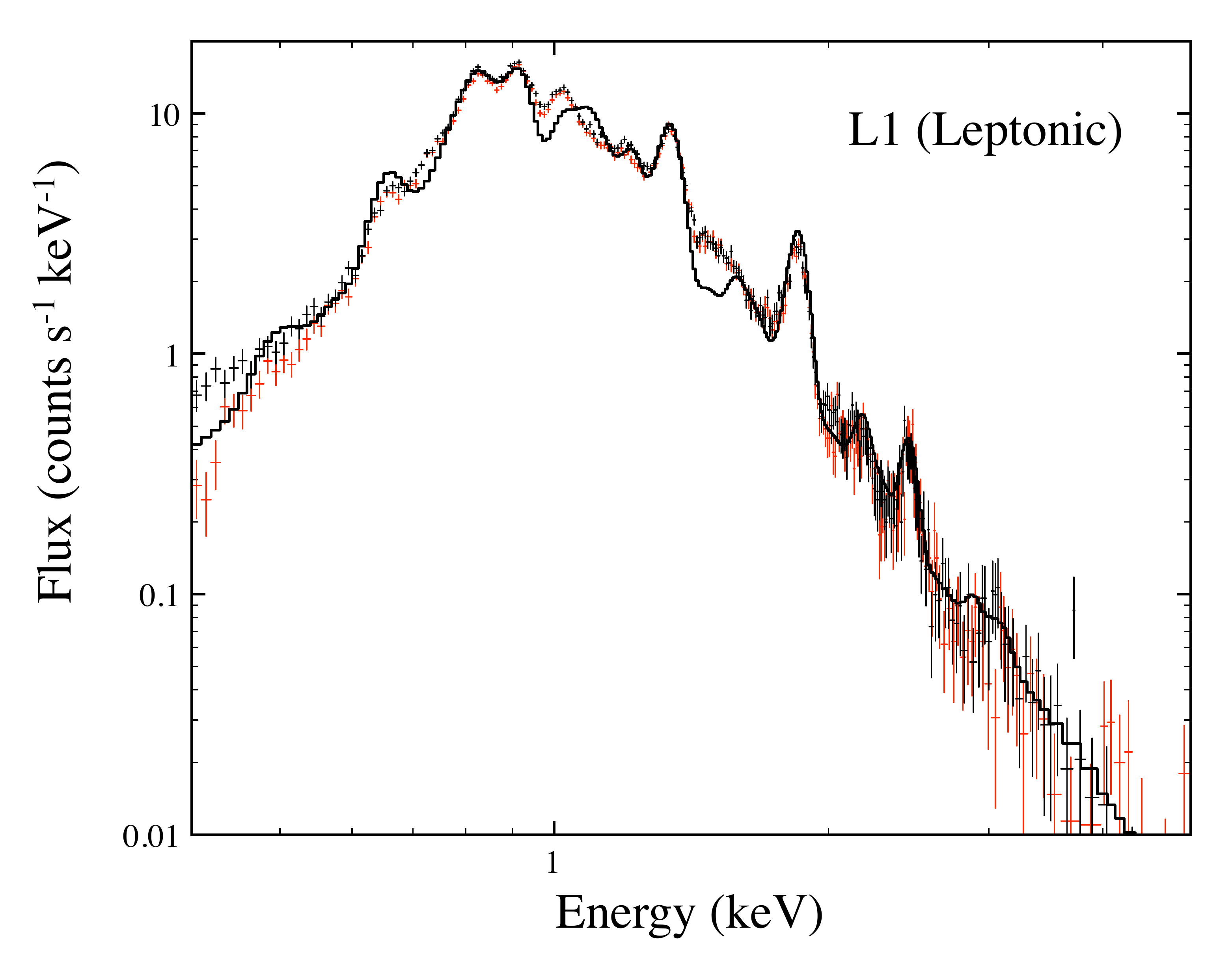}
\includegraphics[width=0.32\textwidth]{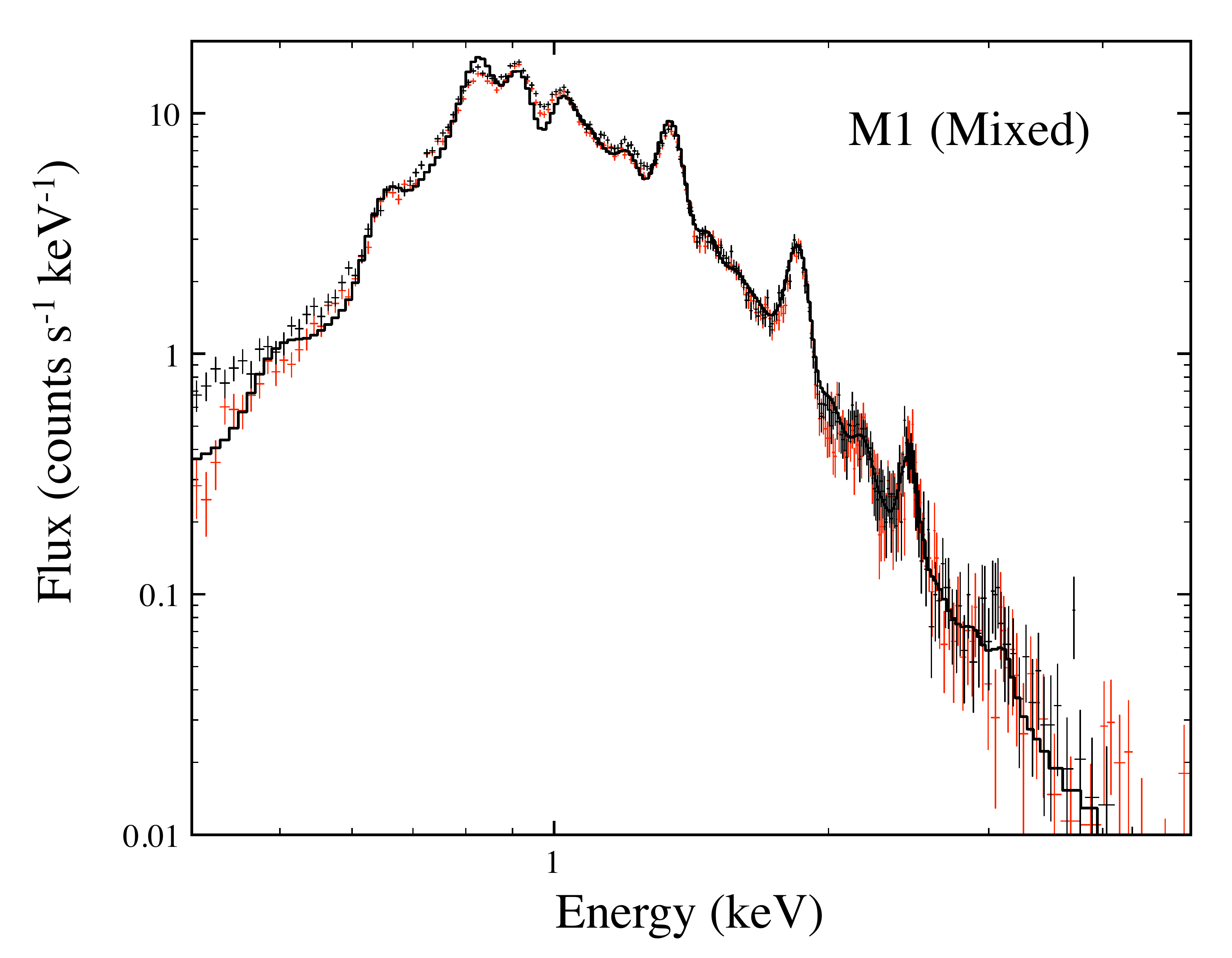}
\caption{The {\it top row} shows broadband fits to radio \citep[open squares,][]{Kothes2006}, and {\it Fermi}-LAT (black circles) observations of \snr\ with the hadronic ({\it top left}), leptonic ({\it top middle}), and mixed ({\it top right}) models, all with uniform CSM density profiles. The blue triangle and horizontal bar represent the total X-ray flux in the 0.5-5.0 keV range estimated from the {\scshape sedov} model fit to the \xmm\ \mosiii\ data. The solid black line is the total nonthermal emission predicted by the model. The modeled spectra from inverse Compton emission (dashed), $\pi^0$-decay (dotted), and nonthermal bremsstrahlung (dot-dashed), are also shown. The {\it bottom row} presents the \xmm\ EPIC MOS 1/2 background subtracted spectra of the region illustrated in Figure \ref{fig:XMMmap} of \snr\ as fit by the model. The histograms in the {\it bottom left}, {\it bottom middle}, and {\it bottom right} panels represent the model X-ray spectral energy distributions from the hadronic, leptonic, and mixed scenarios respectively. See the text for discussion and Tables \ref{tab:crhydro} and \ref{tab:crh} for model parameters.}
\label{fig:allspec}
\end{center}
\end{figure*}

For the hadronic scenario the relativistic electron to relativistic proton ratio was set to $K_{ep}=0.01$, which is consistent with observations of cosmic ray abundances at Earth \citep{Hillas2005}. No combination of ambient proton density and age was found to simultaneously fit the measured radius $R_{\text{SNR}}=(16.0\pm0.5)d_3$ pc at $d_3=1$, and the broadband spectrum observed. Instead, the hadronic scenario discussed here was obtained using $n_0=1.1$ cm$^{-3}$,  and  $t_{\text{SNR}}=15000$ years, which results in an FS radius $R_{\text{FS}}\simeq14.2$ pc. In order to match the observed angular size of \snr\ the distance must hence be $D_{\text{SNR}}=2.7$ kpc in this model. In such case $B_{\text{amp}}=7.0$, and $\varepsilon_{\text{CR}}=23\%$ are required to fit the radio and $\gamma$-ray spectral data. The resulting FS speed is $V_{\text{FS}}\simeq370$ km~s$^{-1}$, the total CSM mass swept up by the half shell is approximately 470 $M_{\odot}$, the overall compression ratio is $R_{\text{tot}}\simeq4.4$, the magnetic field immediately behind the shock is $B_2\simeq67$~$\mu$G  and the total fraction of the SN explosion energy deposited in CR ions $\simeq0.40$. The broadband nonthermal spectrum from this model is compared to the observations in Figure \ref{fig:allspec} ({\it top left panel}), and the parameters are summarized in Table \ref{tab:crh}. The total X-ray flux in the range 0.5-5.0 keV, estimated from the {\scshape sedov} model fit to the \xmm\ \mosiii\ data, is also shown in the figure. The ratio of the pion decay flux to the inverse Compton flux in the energy range 100 MeV to 1 TeV in this scenario is $F_{\pi}/F_{\text{IC}}\simeq14$.

\begin{table*}
\begin{center}
\begin{threeparttable} 
\caption{CR-Hydro-NEI Parameters}
\label{tab:crh}
\begin{tabular}{lccccccccccccc}
\toprule
\noalign{\smallskip}
&&&$n_0$&$D_{\text{SNR}}$&$t_{\text{SNR}}$&&$\varepsilon_{\text{CR}}$&$R_{\text{FS}}$&&$B_2$\tablenotemark{a} \\
\noalign{\smallskip}
Model&$K_{ep}$&s&(cm$^{-3}$)&(kpc)&($10^3$yr)&$B_{\text{amp}}$&(\%)&(pc)&$R_{\text{tot}}$&($\mu$G)&$E_{\text{CR}}$/$E_{\text{SN}}$& $F_{\pi}/F_{\text{IC}}$\tablenotemark{b}\\
\noalign{\smallskip}
\midrule
\noalign{\smallskip}
H1 (Hadronic)&0.01&0&1.1&2.7&15&7.0&23&14.2&4.4&67&0.40&14\\
\noalign{\smallskip}
L1 (Leptonic)&0.02&0&0.2&3.0&8.5&2.7&33&15.6&4.7&26&0.42&0.35\\
\noalign{\smallskip}
M1 (Mixed)&0.015&0&0.5&2.8&11&2.6&35&14.6&4.7&26&0.49&1.0\\
\noalign{\smallskip}
\noalign{\smallskip}
\bottomrule
\noalign{\smallskip}
\end{tabular}
\begin{tablenotes}[para]
\item[a]{$B_2$ is the magnetic field immediately behind the shock.}

\item[b]{$F_{\pi}/F_{\text{IC}}$ is the ratio of the pion decay flux to the inverse Compton flux in the energy range 100 MeV to 1 TeV.}

\end{tablenotes}
\end{threeparttable} 
\end{center}
\end{table*}

The observed X-ray spectrum from \snr\ appears to be almost completely thermal in nature, as shown in Section 2.2, and the expected synchrotron flux in the X-ray band from the model adapts well to this fact. The thermal X-ray emission model obtained from the plasma emissivity code was combined with the non-thermal emission in the 0.4-5.0 keV range to obtain the complete X-ray emission profile. Figure \ref{fig:allspec} ({\it bottom left panel}) compares the resulting spectrum model, folded through the instrument responses of the \mosiii\ instruments, to the \xmm\ observations. The {\scshape tbabs} prescription of photon absorption by the ISM has been used as a multiplicative model on {\scshape xspec} as the only free component of the model fit to the data (\nH$\simeq 0.85\times 10^{22} \text{cm}^{-2}$). 

As shown in Figure 5 ({\it bottom left panel}), the computed thermal X-ray emission from the CR-Hydro-NEI model reproduces the observed X-ray spectrum, with some notable and important exceptions: (1) the modeled X-ray spectrum does not include the clearly visible emission feature at $\sim$1.2 keV;  and (2) the Ly-$\alpha$ emission from Mg, Si, and S appears to be overpredicted relative to the data. The model does, however, reproduce the soft-band emission below 900 eV as well as the continuum shape. Additionally, the modeled flux is consistent with that which was modeled in the {\scshape sedov} model. 

The feature at 1.2 keV is likely associated with Fe-L shell emission. At these energies, the atomic data for iron (specifically \ion{Fe}{17} -- \ion{Fe}{24})  in the emissivity code are incomplete \citep{Patnaude2010}. To compensate for this feature, we include in the fits a Gaussian component at this energy.
The large predicted ratio of the Lyman-$\alpha$ emission to the He-like states is endemic to the model: the hadronic model requires a high density in order to reproduce the observed GeV--TeV emission. In the swept-up shocked material, this results in more collisions per ion per unit time, which eventually leads to higher overall charge states.  The models overpredict the flux from Ly$\alpha$ lines relative to the triplet states in the hadronic model shown in Figure 5 ({\it bottom left panel}). 


Thus we conclude that hadronic processes resulting from the acceleration of particles in \snr\ are a possible origin of the $\gamma$-ray emission observed with the \fermi, but a distance to the SNR on the lower end of the latest estimates would be required.

In the {\it top middle panel} of Figure 5, we show the best fit leptonic model. In order to find a leptonic dominated scenario which fit the broadband spectral properties as well as the the size of \snr, the relativistic electron to relativistic proton ratio was required to be $K_{ep}=0.02$. 
We note that this value is consistent with that found by \citet{Ellison2010} for SNR RX J1713.7-3946. In such case, $t_{\text{SNR}}=8500$ years, $n_0=0.2$ cm$^{-3}$, $B_{\text{amp}}=2.7$, and $\varepsilon_{\text{CR}}=33\%$ are required to fit the radio and $\gamma$-ray spectral data, and the observed radius. At the end of the simulation the radius of the FS is $R_{\text{FS}}\simeq15.6$ pc, the FS speed is $V_{\text{FS}}\simeq780$ km~s$^{-1}$, the total CSM mass swept up by the half shell  is approximately 110 $M_{\odot}$, the overall compression ratio is $R_{\text{tot}}\simeq4.7$, the magnetic field immediately behind the shock is $B_2\simeq26$~$\mu$G  and the total fraction of the SN explosion energy deposited in CR ions $\simeq0.42$, and $F_{\pi}/F_{\text{IC}}\simeq0.35$. The resulting broadband nonthermal spectrum is compared to the observations in Figure \ref{fig:allspec} ({\it top middle panel}).

The expected synchrotron flux in the X-ray band from the leptonic dominated model is also weak in comparison to the \xmm\ observation, and the thermal X-ray emission model obtained from the plasma emissivity code, shown in Figure \ref{fig:allspec} ({\it bottom middle panel}), fits the observed spectrum reasonably well. Much like the hadronic scenario, the thermal X-ray emission fits the data qualitatively (the absorbed flux in the range 0.5-5.0 keV predicted $6.7\times10^{-11}$erg cm$^{-2}$ s$^{-1}$, also matches the flux extracted with the {\scshape sedov} model), with notable exceptions again related to the flux ratio of He-like to Ly$\alpha$ emission. In the leptonic scenario, the emission from Ly$\alpha$ states is underpredicted. The reason here is that in the lower density leptonic scenario, the shocked plasma is not ionized to as high a degree as in the hadronic model.
This model provides a reasonable overall description of the observed nonthermal broadband spectrum, thermal X-ray emission, and radial extent of \snr. Thus, nonthermal radiation from shock accelerated leptons may be the dominant source of the $\gamma$-ray emission observed in the direction of SNR \snr.

Finally, we consider a ``mixed'' scenario, shown in the {\it top right panel} of Figure 5. The parameters are shown in Table \ref{tab:crh}, and the broadband and X-ray spectral models illustrated in the {\it middle top} and {\it middle bottom} panels respectively. The key parameters in this model are the $K_{ep}=0.015$, $t_{\text{SNR}}=11,000$ years, $n_0=0.5$ cm$^{-3}$, $B_{\text{amp}}=2.6$, and $\varepsilon_{\text{CR}}=35\%$. This input set results in a radius $R_{\text{FS}}\simeq14.6$ pc, FS speed $V_{\text{FS}}\simeq520$ km~s$^{-1}$, total CSM mass swept up by the half shell $M_{\text{sw}}\approx$ 220 $M_{\odot}$,  overall compression ratio $R_{\text{tot}}\simeq4.7$, magnetic field immediately behind the shock $B_2\simeq26$~$\mu$G  and $E_{\text{CR}}$/$E_{\text{SN}}\simeq0.49$. In terms of the thermal X-ray emission, this model appears to provide the best fit. The relative flux in the modeled He-like and Ly$\alpha$ lines is consistent with the observations (in contrast to both the hadronic and leptonic scenarios), and the modeled emission around $\sim$ 1 keV also agrees well with the data, both in overall normalization as well as which lines are present (notably, the shape of the spectrum around 1 keV in the leptonic scenario does not agree with the data, while in the hadronic model, where the shape is consistent with the data, the model consistently underpredicts the flux). 

While all the models fit the broadband nonthermal spectrum, the inclusion of the thermal X-ray emission allows us to rule out certain models. In particular, the thermal X-ray emission favors a model where $F_{\pi}/F_{\mathrm{IC}}$ $\approx$ 1. In this case, we find that we can reproduce both the nonthermal emission as well as the shape and bulk properties of the thermal X-ray emission.

Since \citet{Kothes2012} considered all other previous distance estimates and, using new data analysis, concluded that the remnant is in the Perseus Arm, at a distance of approximately 3 kpc, we do not focus on a full treatment of the $E_{\text{SN}}= 7\times 10^{51}$ erg, $D_{\text{SNR}} = 7.5 $ kpc scenario proposed by \citet{Durant2006}. However, we did explore this set of parameters to establish whether new constraints on the distance could be determined by modeling the broadband characteristics. We found that we could not rule out the larger-distance parameter set, and that the main constraint that can be drawn is that the gamma-ray emission in such scenario must be the result of inverse Compton emission, since hadronic and mixed cases cannot fit the SNR characteristics appropriately.

\subsection{Other possible sources of $\gamma$-rays}

There are some alternative scenarios explaining the $\gamma$-ray emission detected that should be considered. One possibility is that the $\gamma$-rays result of high energy radiation from AXP \ps. Another scenario is that where the particles accelerated at the SNR shock are interacting with the dense molecular material nearby. 

The AXP \ps\ has been carefully studied in the X-ray band since it was first detected by \citet{Gregory1980}. Observations with the {\it Ginga} satellite found a pulsar period of $P\simeq7$ s, and spin-down rate $\dot{P}\simeq(3-6)\times10^{13}$ s s$^{-1}$ \citep{Iwasawa1992}. In 2002, multiple outbursts similar to those of soft-gamma repeaters (SGRs) were detected from \ps, and changes in the pulsed flux, persistent flux, energy spectrum, pulse profile, and spin-down of the source were also evidenced. 
 
The \fermi\ has detected a large population of $\gamma$-ray pulsars \citep{Abdo2010a,Abdo2011b}, yet a recent search for $\gamma$-ray emission from magnetars with this instrument was unsuccessful \citep{Abdo2010b}. These authors found no definitive evidence of a source coincident with \ps\, and they derived a flux upper limit in the 0.1-10 GeV range of 1.7$\times 10^{-11}$ erg s$^{-1}$ cm$^{-2}$, with TS value 15.6. We do not find a $\gamma$-ray source spatially coincident with \ps\ either. The local maximum of the TS is not coincident with the magnetar position, and it is located at approximately 0.1$^{\circ}$ SE of the AXP. Hence, while the AXP cannot be ruled out as the source of the $\gamma$-ray flux observed, it appears unlikely due to the spatial characteristics of the emission. 

Since the SNR is embedded in a complex region with large areas of dense molecular material, these clouds represent an effective target for interacting with CRs accelerated at the SNR shock and generating $\gamma$-rays. The two regions that appear closest to the remnant are the large cloud which halted the expansion of the SNR on the west, and the cloud associated with the X-ray bright feature called the ``Lobe'' \citep[and references therein]{Sasaki2004}, whose position is indicated in Figure \ref{fig:XMMmap}. 

\begin{figure}
\begin{center}
\includegraphics[width=\columnwidth]{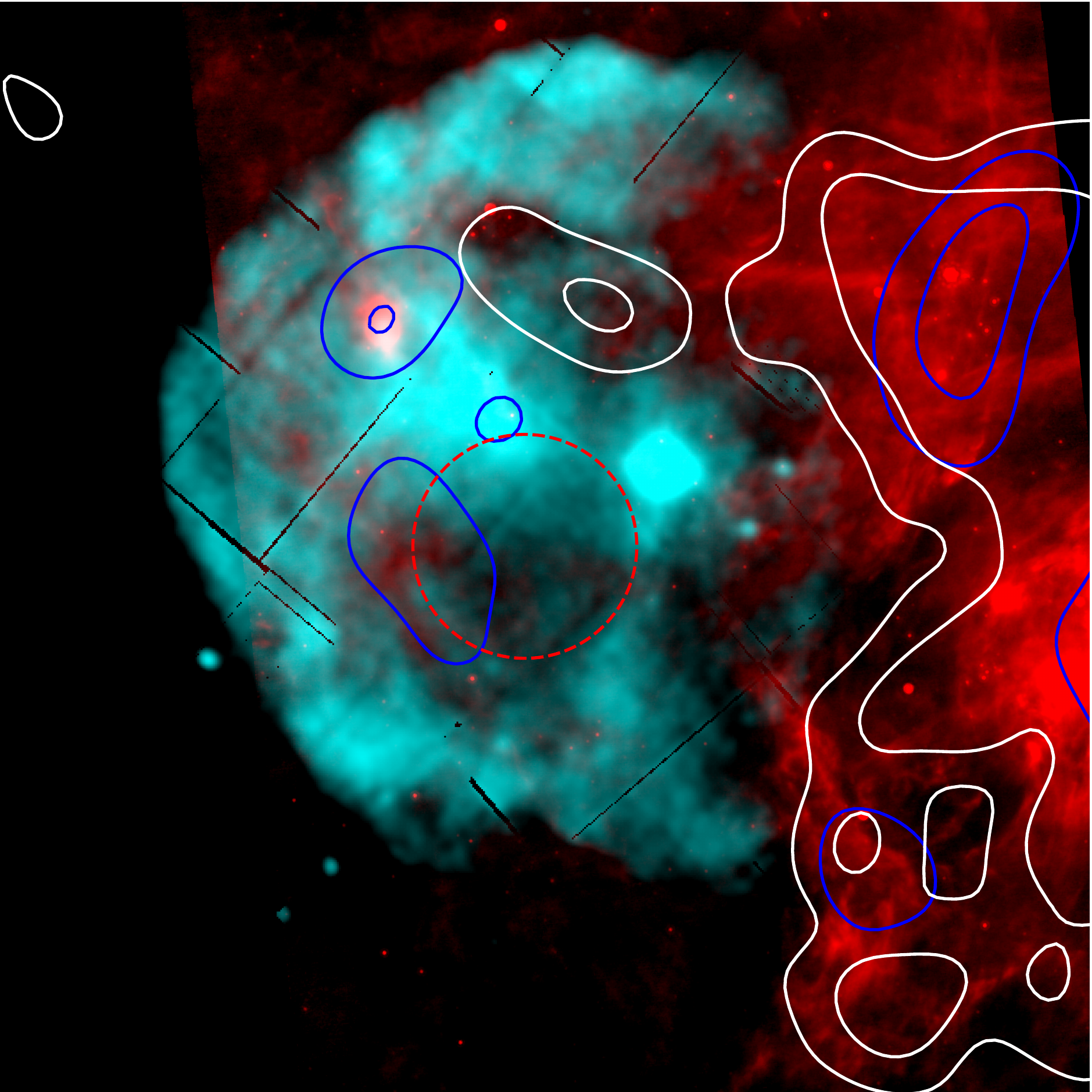}
\caption{Combined multi-wavelength view of the region around \snr. The smoothed, merged \xmm\ MOS 1/2 intensity map in the range 0.5-5.0 keV is shown in cyan, while the {\it Spitzer} MIPS 24$\mu$m image is shown in red. The white (blue) contours correspond to the CO line emission as observed by the CGPS, at velocity -51 (-54) km/s, and the levels correspond to 0.5 and 1.5 K. The 95\% confidence radius of the centroid of the {\it Fermi}-LAT source is shown as a dashed red circle }
\label{fig:multi}
\end{center}
\end{figure}

The scenario in which the $\gamma$-rays result from CRs interacting with the larger cloud to the west of \snr\ is clearly ruled out by the spatial extent and location of the $\gamma$-ray emission detected, which is concentrated on the eastern part of \snr. The relative positions of the western molecular cloud and the $\gamma$-ray emitting region are shown in Figure \ref{fig:multi}, which combines the \xmm\ intensity map with {\it Spitzer} MIPS 24$\mu$m data, and CO line emission contours from observations of the Canadian Galactic Plane Survey \citep{Kothes2002}. In this figure the infrared and CO line emission (at velocity -51 km/s) suggest that the GMC is mostly confined to the western part of the field, while the $\gamma$-ray emission is located east of the AXP (the 95\% confidence radius of the centroid is shown as a dashed red circle). 

A fraction of the CRs accelerated at the shock of an SNR are expected to escape and diffuse away, and their interaction with nearby regions of dense material will produce $\gamma$-rays \citep[][and references therein]{Ellison2011}. Since no significant emission from the western cloud is detected by the Fermi-LAT, we considered what constraints this places on the nature of particle escape from the acceleration site. For this purpose we used CR-Hydro-NEI and conservative estimates of the values for the mass of the cloud ($M_{\text{cloud}}\sim10^3M_{\odot}$), the density of the cloud material ($n_{\text{cloud}}\sim150$ cm$^{-3}$), its distance to the shock ($l\sim0.2$ pc) \citep{Tatematsu1987,Tatematsu1990}, and the diffusion coefficient for CRs in this region \citep[$D_{\text{ISM}}\sim10^{26}$ cm$^2$ s$^{-1}$,][]{Fujita2010}. For each of the three parameter sets used in Section 3.3 the simulation results in $\sim3-5\%$ of the explosion energy being placed in escaping particles. Based on this, we estimate that the resulting $\gamma$-ray flux from the Western cloud would hence be $\lesssim20\%$ of the observed flux from \snr, which is below the sensitivity level of the {\it Fermi}-LAT. We conclude that a significantly larger population of escaping cosmic rays, or an extremely high mass for the western cloud, would be required to produce a detectable $\gamma$-ray signature from this region.

\citet{Sasaki2006} used CO and X-ray observations with the {\it Chandra} X-ray Telescope to observe the Lobe, and concluded that there was clear evidence that the SNR shock was interacting with the molecular cloud complex at that position. These authors found three CO clouds near the Lobe with radial velocities between -52 and -57 km/s, which can be clearly seen as blue contours on Figure \ref{fig:multi}. The brighter cloud, located to the NE of the X-ray Lobe, is also clearly coincident with a bright infrared region (as shown in Figure \ref{fig:multi}), IRAS 23004+5841, which \citet{Woo1987} established has the IR colors of a star-forming region. \citet{Sasaki2006} believe the X-ray Lobe is the result of the partial evaporation of the eastern cloud, and estimate the cloud properties through analysis of the X-ray emission. Assuming the shocked cloud material fills a sphere of radius 3$'$, and a distance of 3 kpc, they derive a density of $n_{\text{Lobe}}=0.9$ cm$^{-3}$. 

If the $\gamma$-rays originate from the proton-proton interactions between cosmic rays accelerated at the shock of \snr\ and the shocked material in the Lobe, one can derive the density of the shocked cloud by considering the $\gamma$-ray flux observed. \citet{Drury1994} derive the pion decay flux above 100 MeV to be $F_{100\text{ MeV}}\approx 4.4\times10^{-7} (\theta\, E_{\text{51}} \,n_{\text{cm}^{-3}}\, D^{-2}_{\text{kpc}}$) cm$^{-2}$ s$^{-1}$, where $E_{\text{51}}$ is the SN explosion energy in units of 10$^{51}$ ergs, the fraction of the explosion energy deposited in accelerated particles is represented by  $\theta$,  the density of the surrounding material $n_{0}$ is in units of cm$^{-3}$, and $D_{\text{kpc}}$ is the distance to the SNR in kpc. Using reasonable values for the parameters such as $E_{\text{51}}=1$, $\theta=0.4$, and $D_{\text{kpc}}=3$, as well as assuming that cosmic ray production in the SNR shock occurs isotropically, and that the fraction of the accelerated particles incident on the Lobe material is equal to the ratio of the angular area of the cloud to that of the entire SNR shell (presuming that the CRs trapped in the shock interact only with the portion of the cloud that has been overtaken by the shock), the density of the Lobe required to produce the observed $\gamma$-ray flux (calculated in Section 2.1) is $n_{\text{Lobe}}\approx120$ cm$^{-3}$. Hence, there is a large discrepancy between the observational estimate of the density of the Lobe from the X-ray emission ($n_{\text{Lobe}}=0.9$ cm$^{-3}$) obtained by \citet{Sasaki2006}, and the density suggested by combining the analysis of \citet{Drury1994} and the observed $\gamma$-ray flux.

While the Lobe density estimates obtained from the X-ray and $\gamma$-ray fluxes differ by approximately two orders of magnitude, it is not possible to rule out the Lobe as the origin of the emission above 100 MeV. Instabilities in the postshock flow upon interacting with dense ambient media may result in considerable clumping of the molecular material. The shocks propagating through clumps of dense material will then be expected to quickly become radiative and their X-ray yields will become much less significant, resulting in density underestimates. The $\gamma$-ray flux from the interaction of cosmic rays with the shocked cloud is not expected to be affected by clumping, and hence high filling factors of clumped material may be sufficient to explain the ratio of $\gamma$-ray to X-ray flux. This scenario has been suggested to explain discrepancies between the X-ray and $\gamma$-ray characteristics of several SNRs, including RX J1713.7-3946 \citep{Inoue2012}, and 3C391 \citep{Castro2010}.

\section{Conclusions}

X-ray and radio observations of \snr\ reveal this SNR to be a semi-spherical shell, likely due to interaction with a giant molecular cloud in the west which halted the remnant's expansion in that direction. The X-ray spectrum of the SNR is dominated by thermal emission, and no signature of non thermal radiation from the shell in this energy band has been detected. Additionally, \snr\ contains AXP \ps\ which is thought to be the compact remnant of the progenitor core-collapse supernova explosion. 

We report the detection with the {\it Fermi}-LAT of a possibly extended $\gamma$-ray source coincident with SNR \snr, and use the broadband properties of the remnant to constrain the origin of this MeV-GeV emission. We use a model of SNR evolution and emission that includes hydrodynamics, efficient cosmic ray acceleration, nonthermal emission mechanisms and a self-consistent calculation of the X-ray thermal emission in order to consider the system's parameters required for the SNR to be the origin of the LAT source. 
We find that the broadband radio, X-ray, and gamma-ray observations of \snr\ can be reasonably fit with two distinct parameter sets where the dominant emission mechanism behind the gamma-ray flux is either hadronic (pion-decay) or leptonic (inverse Compton) in nature. However, the inclusion of the thermal X-ray emission in the fitting results in stronger constraints on the origin of the $\gamma$-ray flux and indicates that an intermediate scenario, where both relativistic leptons and hadrons contribute significantly to the MeV-GeV emission, provides a much superior fit to the data than the extreme cases.

Cosmic rays interacting with high-density media are expected to result in $\gamma$-ray emission from relativistic ions colliding with ambient hadrons. Hence, we consider the interaction between \snr\ and nearby molecular clouds as a possible origin of the {\it Fermi}-LAT source, and while it is possible to rule out an association with the GMC located to the west because it is too far from the centroid of the $\gamma$-ray emission, we find that it is possible that the source is related to an interaction between the SNR and a region of shocked molecular material known as the X-ray ``Lobe''. Additionally, we find the association between the $\gamma$-ray source and the AXP \ps\ to be unlikely due to spatial considerations.

\acknowledgments
The authors thank Yasunobu Uchiyama, Stefan Funk, Joshua Lande, Elizabeth Hays and Marianne Lemoine-Goumard for their advise on this work, especially with regard to {\it Fermi}-LAT data analysis. Additionally, we thank the referee for his/her feedback. This work was partially funded by NASA Fermi grant NNX10AP70G. 

DC also acknowledges support for this work provided by the National Aeronautics and Space Administration through the Smithsonian Astrophysical Observatory contract SV3-73016 to MIT for Support of the Chandra X-Ray Center, which is operated by the Smithsonian Astrophysical Observatory for and on behalf of the National Aeronautics Space Administration under contract NAS8-03060. POS acknowledges partial support from NASA contract NAS8-03060. DCE acknowledges support from NASA ATP grant NNX11AE03G. 

\bibliographystyle{apj}

\end{document}